\documentclass[aps,prb,twocolumn,superscriptaddress,showpacs,floatfix]{revtex4}

\usepackage{graphicx,color}
\usepackage{dcolumn}
\usepackage{bm}
\usepackage{stmaryrd}
\usepackage{latexsym}
\usepackage{amssymb}
\usepackage{amsfonts}
\usepackage{amsmath}
\usepackage{epstopdf}

\begin{document}

\title{Impurity- and Magnetic-field-induced Quasiparticle States in Chiral $p$-wave Superconductors}

\author{Yao-Wu Guo}
\affiliation{Department of Physics and State Key Laboratory of Surface Physics, Fudan University, Shanghai 200433, China}
%\affiliation{Collaborative Innovation Center of Advanced Microstructures, Nanjing 210093, China}

\author{Wei Li}
\email{liwei@mail.sim.ac.cn}
\affiliation{State Key Laboratory of Functional Materials for Informatics and Shanghai Center for Superconductivity, Shanghai Institute of Microsystem and Information Technology, Chinese Academy of Sciences, Shanghai 200050, China}
\affiliation{CAS Center for Excellence in Superconducting Electronics, Shanghai 200050, China}

\author{Yan Chen}
\email{yanchen99@fudan.edu.cn}
\affiliation{Department of Physics and State Key Laboratory of Surface Physics, Fudan University, Shanghai 200433, China}
\affiliation{Collaborative Innovation Center of Advanced Microstructures, Nanjing 210093, China}

\date{\today}


\begin{abstract}
Both impurity- and magnetic-field-induced quasiparticle states in chiral $p$-wave superconductors are investigated theoretically by solving the Bogoliubov--de Gennes equations self-consistently. At the strong scattering limit, we find that a universal state bound to the impurity can be induced for both a single nonmagnetic impurity and a single magnetic impurity. Furthermore, we find that different chiral order parameters and the corresponding supercurrents have uniform distributions around linear impurities. Calculations of the local density of states in the presence of an external magnetic field show that the intensity peak of the zero-energy Majorana mode in the vortex core can be enhanced dramatically by tuning the strength of the external magnetic field or pairing interaction.

\vspace{5mm} \noindent {\bf Keywords:} nonmagnetic/magnetic impurity, chiral \emph{p}-wave superconductor, vortex state, Majorana mode

\end{abstract}

\maketitle

\section{Introduction}
Topological aspects of Fermi systems in solids and ultracold atoms~\cite{WXG,Nayak,ZHSC,Kane,A.Kitaev,Xiangtao}, such as quantum Hall effects~\cite{hall effect,hall effect2,hall effect3,hall effect4}, topological insulation, and topological superconductivity~\cite{TI, TI2}, have attracted much attention over the past three decades. Moreover, a similar topologically nontrivial ground state has been found in the superfluid A-phase of $^3$He films \cite{He3} with chiral $p_{x}+ip_{y}$ symmetry and in the layered triplet $p$-wave superconductor Sr$_2$RuO$_4$ \cite{srruo}, which is expected to host Majorana modes~\cite{majorana, p-wave}. Previous theories have proposed that the zero-energy Majorana bound state can be realized in the vortex core of spinless $p_{x}+ip_{y}$-wave superconductors or superfluids~\cite{p-wave,superfluid}. The zero-energy fermionic modes can be described in terms of self-conjugated Majorana modes, which are also expected to occur in other systems, such as the $\nu=\frac{5}{2}$ fractional quantum Hall state~\cite{qhe,MGreiter} and the surface state of three-dimensional topological insulators with proximity coupling to conventional $s$-wave superconductors~\cite{surface1, surface2, surface3}. One of the simplest effective models realizing a topological superconducting phase supporting Majorana modes is the two-dimensional (2D) chiral $p_{x}+ip_{y}$-wave superconductor. We focus on the impurity effects and the vortex core state structure of a 2D chiral $p$-wave superconductor, identifying its topological nature and exploring its local physics.

This paper theoretically investigates the interplay of the ground-state topologies and properties of fermionic bound states near impurities and topological defects. The universal bound state of the quasiparticle and the supercurrent induced by a single impurity in a chiral $p$-wave superconductor at a sufficiently high scattering strength can be observed for both a single nonmagnetic impurity and a single magnetic impurity. For a row of linear nonmagnetic impurities and a row of linear magnetic impurities in a chiral $p$-wave superconductor, we find that the distributions of the supercurrent and chiral domain structures of the $p_{x}\pm ip_{y}$-wave order parameters are different, because the former system preserves time-reversal symmetry, whereas the latter breaks it. Additionally, the directions and magnitudes of the supercurrent have different distributions for two degenerate $p_x\pm ip_y$-wave order parameters. This finding may provide a route to distinguishing the two degenerate components of a chiral $p$-wave superconductor. Further, we study a topological defect in the vortex core structure of a chiral $p$-wave superconductor and show that the local density of states (LDOS) at the vortex core center has a zero-energy peak, which may correspond to the Majorana mode~\cite{17}. All these theoretical findings
have potential applications in experimental explorations of Majorana modes.

The remainder of the paper is organized as follows. Section~\ref{sec:bdg} discusses primarily the theoretical model Hamiltonian and the methods of numerical calculation. Section~\ref{sec:singleimpurity} details numerical results for both a single nonmagnetic impurity and a single magnetic impurity as well as the effects of linear impurities. Section~\ref{sec:splitting} considers the splitting of the zero-energy peak of the vortex core states in a chiral $p$-wave superconductor in the presence of an external magnetic field. Finally, conclusions drawn from the main results of the study are presented in Section~\ref{sec:conclusion}.

\section{Model and Method}
\label{sec:bdg}

In the following study, we adopt an effective Bardeen--Cooper-- Schrieffer (BCS)-type mean-field Hamiltonian defined on a 2D triangular lattice:
\begin{eqnarray}
\hat{H}_{\text{eff}}&=&-\sum_{\langle i,j\rangle \sigma}(t_{ij}\hat{c}_{i\sigma }^{\dagger }\hat{c}_{j\sigma}
+\text{h.c.})+\sum_{i,\sigma }(\emph{V}_{i\sigma}^{imp}-\mu) \hat{c}_{i\sigma }^{\dagger }\hat{c}_{i\sigma } \nonumber\\%
&+&\sum_{\langle i,j\rangle} \left [\Delta
_{ij}^{\pm}(\hat{c}_{i\uparrow }^{\dagger }\hat{c}_{j\downarrow }^{\dagger}\pm
\hat{c}_{i\downarrow }^{\dagger }\hat{c}_{j\uparrow
}^{\dagger})+\text{h.c.}\right ],\label{Eq1}
\end{eqnarray}%
where $n_{i\sigma}=\langle \hat{c}_{i\sigma}^{\dagger}\hat{c}_{i\sigma}\rangle$ is the density of electrons with spin $\sigma$ at site $i$, $\mu$ denotes the chemical potential, and $\pm$ represents the spin triplet and singlet pairings, respectively. The pairing potential $\Delta_{ij}^{\pm}$ is defined as $\Delta_{ij}^{\pm}=\frac{V}{2}(\langle \hat{c}_{i\uparrow}\hat{c}_{j\downarrow}\rangle \pm \langle
\hat{c}_{i\downarrow}\hat{c}_{j\uparrow}\rangle)$, which is derived from a mean-field treatment of the pairing interaction $V\sum_{\langle
i,j\rangle}(\hat{c}_{i\uparrow}^{\dagger}\hat{c}_{j\downarrow}^{\dagger}\hat{c}_{i\uparrow}\hat{c}_{j\downarrow}
+\hat{c}_{i\downarrow}^{\dagger}\hat{c}_{j\uparrow }^{\dagger}\hat{c}_{i\downarrow}\hat{c}_{j\uparrow})$. In the presence of an external magnetic field, the hopping integral $t_{ij}$
can be rewritten as $t_{ij}\Rightarrow t_{ij}\exp(i\varphi_{i,j})$ for the
nearest-neighbor sites $\langle i,j\rangle$, where
$\varphi_{i,j}=-\frac{\pi }{\Phi _{0}}\int_{{\bf{r}}_{i}}^{{\bf{r}}_{j}}%
{\bf{A}}({\bf{r}})\cdot d{\bf{r}}$ with ${\bf{A}}({\bf{r}})$ being the
vector potential, and $\Phi _{0}=hc/2e$ being the superconducting flux
quantum. By choosing the Landau gauge, the vector potential ${\bf{A}}({\bf{r}})$ can be rewritten as $(-By,0,0)$, where $B$ is the external magnetic field along the $z$ direction. By performing the Bogoliubov transformation, the BCS-type effective Hamiltonian (\ref{Eq1}) can be diagonalized by solving the Bogoliubov--de Gennes (BdG) equation:
\begin{equation}
\sum_{j}\left(
\begin{array}{cc}
H_{ij,\sigma} & \Delta^{\pm}_{i,j} \\
\mp\Delta _{i,j}^{\pm\ast } & -H_{ij,\bar{\sigma}}^{\ast}
\end{array}%
\right) \left(
\begin{array}{c}
u_{j,\sigma}^{n} \\
v_{j,\bar{\sigma}}^{n}%
\end{array}%
\right) =E_{n}\left(
\begin{array}{c}
u_{i,\sigma}^{n} \\
v_{i,\bar{\sigma}}^{n},
\end{array}%
\right),  \label{BdG}
\end{equation}%
where $u_{i,\sigma}^{n}$ and $v_{i,\bar{\sigma}}^{n}$ are the Bogoliubov quasiparticle amplitudes with corresponding eigenvalue $E_{n}$. $H_{ij,\sigma}=-t_{ij}+\delta_{i,j}(\emph{V}_{ij,\sigma}^{imp}-\mu)$ with $n_{i\sigma}$,
subject to the self-consistent conditions $n_{i\uparrow}=\sum_n|u^n_{i,\uparrow}|^2f(E_n)$ and $n_{i\downarrow}=\sum_n|v^n_{i,\downarrow}|^2[1-f(E_n)]$. Further, $f(E)$ is the Fermi distribution function, and $\Delta_{i,j}$ is calculated as $\Delta^{\pm}_{i,j}=\frac{V}{4}\sum_{n}(u_{i\uparrow}^n
v_{j\downarrow}^{n*}\mp u_{j\uparrow}^n
v_{i\downarrow}^{n*})\tanh(\frac{E_n}{2k_B T}). %
\label{delta}$
We numerically solve the above set of BdG equations (\ref{BdG}) self-consistently. We first guess an initial set of parameters of $\Delta_{ij}$ and $n_i$ and then diagonalize the BdG equations to obtain the eigenvalues $E_n$ and the corresponding eigenfunctions $u_{i\sigma}^{n}$ and $v_{i\sigma}^{n}$ for calculating the new parameters in the next iteration. The calculation is repeated until the difference in the order parameter between two consecutive iteration steps is less than $10^{-4}$. Once self-consistency is achieved, the LDOS can be evaluated using the supercell technique with the summation averaged over $ M_x\times M_y $ wave vectors:
\begin{equation}
\rho_{i}(\omega)=\sum_{n}[|u^{n}_{i\uparrow}|^{2}\delta(E_n-\omega)+
|v^{n}_{i\downarrow}|^{2}\delta(E_n+\omega)],
\end{equation}
where the delta function $\delta(x)$ is taken as $\Gamma/\pi(x^2+\Gamma^2)$ with $\Gamma$ equal to 0.01.

\section{Impurity states}
\label{sec:singleimpurity}

\subsection{Effects of a Single Impurity}

We first present the results of the resonant state induced by a single nonmagnetic impurity or a single magnetic impurity in a chiral $p_x+ip_y$-wave pairing superconductor by self-consistently solving the BdG equation shown in (\ref{BdG}). We choose the lattice constant \emph{a} and hopping integral $t$ as the units of length and energy, respectively. The calculation is performed on a $32\times32$ supercell, and each cell has a size of $32\times32$. We set $\emph{V}=2.4$ with doping $\delta=0.01$ and place a magnetic or nonmagnetic impurity at the center point of $(16,16)$ with an impurity potential of $V_{imp,ij}^{\sigma}=V_{imp}^{\sigma}\delta_{ij}$.
We set $V_{imp}^{\uparrow}=-V_{imp}^{\downarrow}=V_{0}$ for the magnetic impurity and $V_{imp}^{\uparrow}=V_{imp}^{\downarrow}=V_{0}$ for the nonmagnetic impurity.

\begin{figure}[tbp]
\centering
\resizebox{8.4cm}{10cm}{\includegraphics{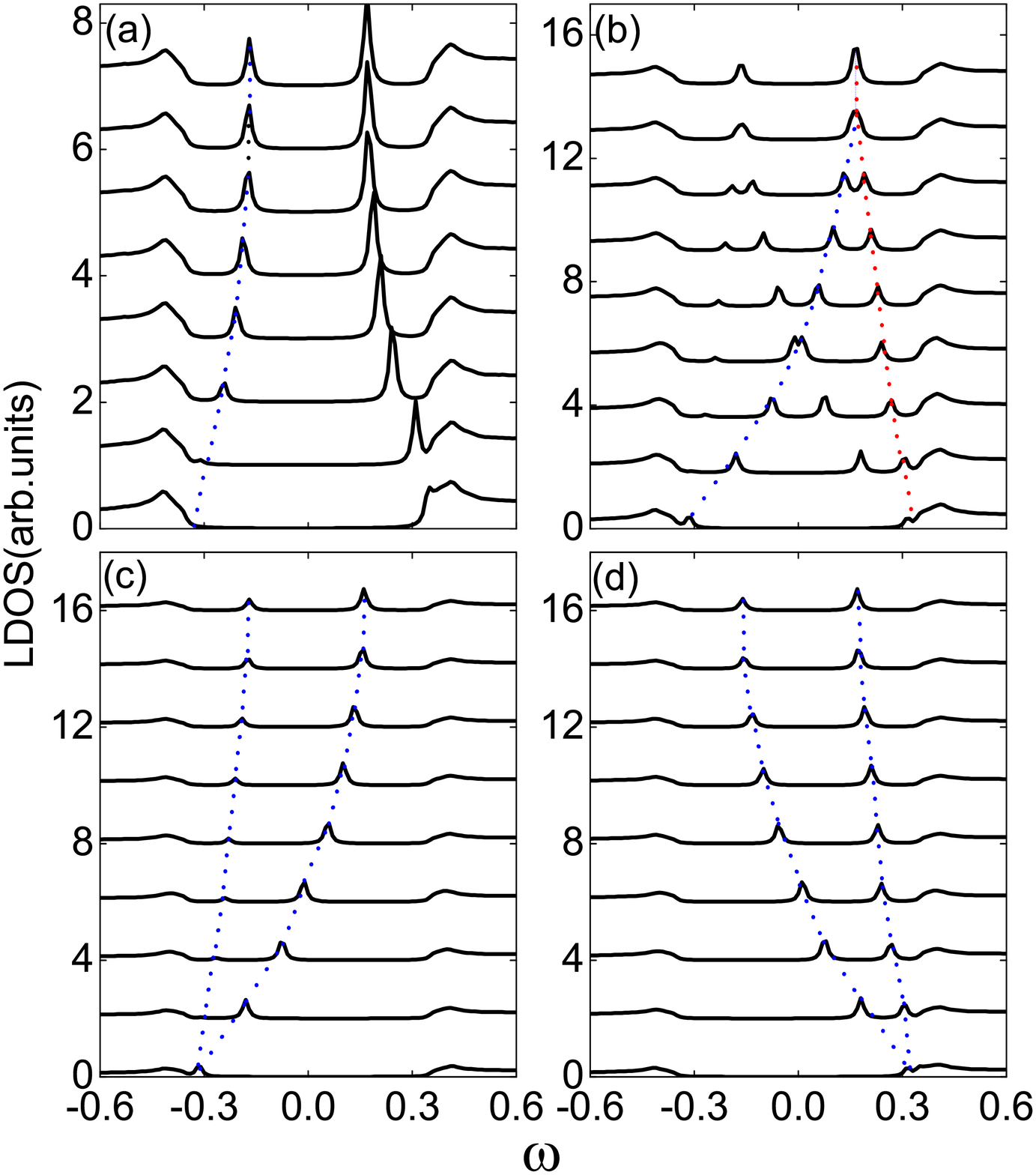}}
\caption{Resonant states induced by (a) a single nonmagnetic impurity and (b) a single magnetic impurity with various scattering
strengths $V_{0}=1,2,5,10,20,50,60,100$ from bottom to top in (a) and with $V_{0}=1,2,3,4,5,10,20,60,100$ in (b). Descriptions of the (c) spin-up species and (d) spin-down species of the resonant state in (b).}
\label{fig:fig1}
\end{figure}

As we know that the density of states in an isotropic $s$-wave superconductor has a fully gapped structure with an energy gap of $\Delta_{0}$, the bound state can appear only at the gap edge~\cite{L.Yu,H.shiba}. However, a single strongly scattering impurity can produce a bound or virtually bound quasiparticle state inside the gap in a $d$-wave superconductor~\cite{A.V.Balatsky,ychen2}. For a chiral $p$-wave superconductor such as Sr$_2$RuO$_4$, the LDOS peaks split near zero energy within the gap, and a different LDOS structure around the impurity may serve as a fingerprint by which to distinguish different pairing states~\cite{M.Takigawa,HHu}.

For a single nonmagnetic impurity [see Fig.~\ref{fig:fig1}(a)], when impurity scattering is switched on, a bound state immediately appears at the edge of the energy gap, in accordance with the result of previous work~\cite{L.Yu,H.shiba}. As the increase in $\emph{V}_{0}$ is gradual, the distance between two bound states at the two gap edges decreases monotonously. At the limit of strong scattering, the energy saturates to $\emph{E}\simeq\Delta_{0}^{2}/\emph{E}_{F}$, where $\Delta_{0}$ is the gap parameter in the absence of an impurity.

\begin{figure}[!t]
\centering
\resizebox{7.6cm}{8.3cm}{\includegraphics{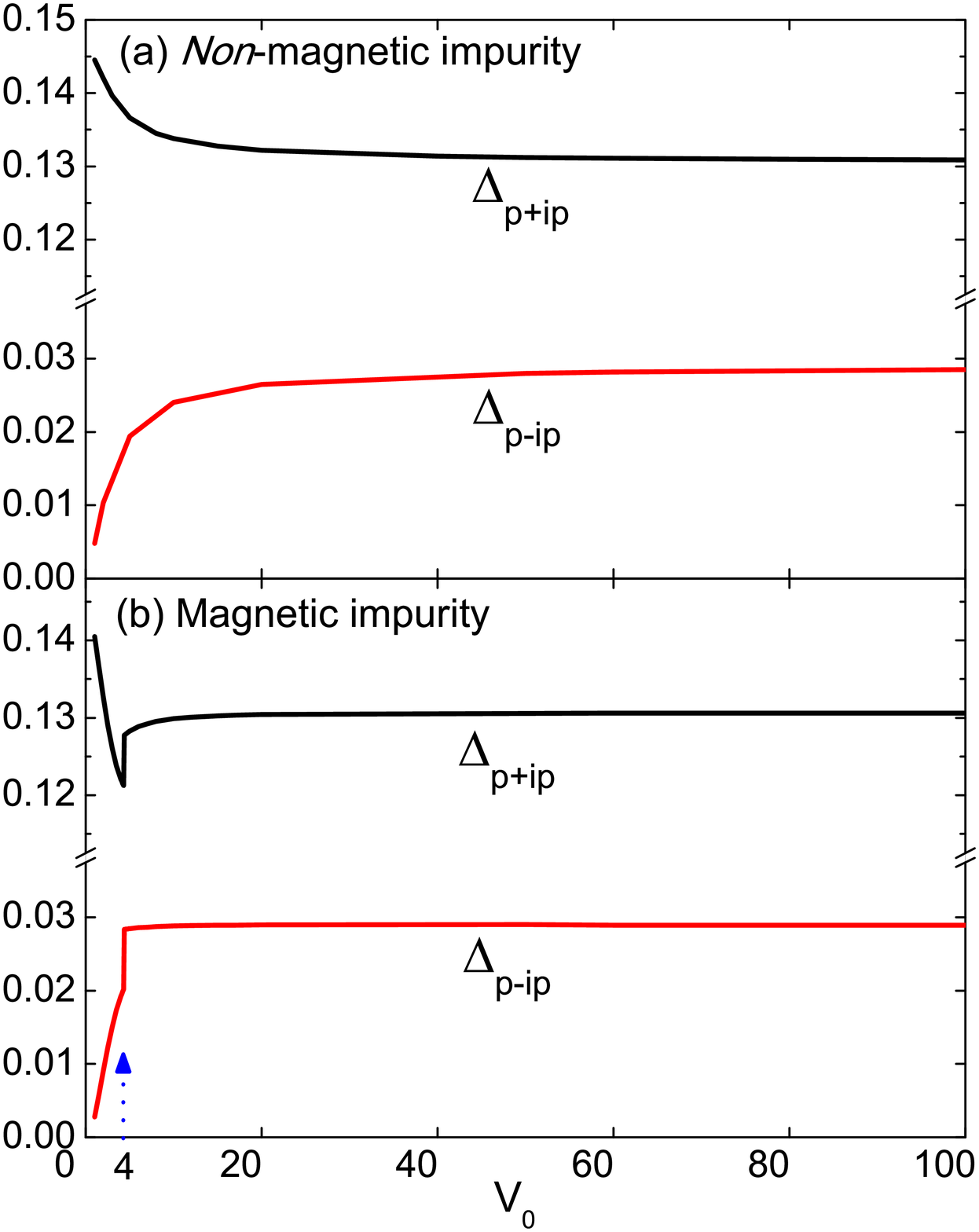}}
\caption{Variation in order parameter values around (a) a
nonmagnetic impurity and (b) a magnetic impurity as a function of the strength of $V_{0}$.}
\label{fig:fig2}
\end{figure}

However, time-reversal symmetry is broken in the presence of a single magnetic impurity; consequently, the system favors a chiral pairing symmetric state, such as a \emph{$p_{x}+ip_{y}$} or \emph{$p_{x}-ip_{y}$} state, which allows the bound states to survive inside the superconducting energy gap [see Fig.~\ref{fig:fig1}(b)]. The spin-resolved LDOS and spin-up and spin-down species of the resonant states are shown in Fig.~\ref{fig:fig1}(c) and (d), respectively. These theoretical calculations clearly suggest that the energy of the bound state is determined by the impurity scattering potential $\emph{V}_{imp}$. As $\emph{V}_{imp}$ increases, the bound state of spin-up electrons moves from the lower energy gap edge toward the upper energy gap edge [see Fig.~\ref{fig:fig1}(c)], whereas the bound state of spin-down electrons moves in the opposite direction [see Fig.~\ref{fig:fig1}(d)]. When the strength of the impurity scattering potential increases to the strong scattering limit, the bound states saturate to $\emph{E}\simeq\Delta_{0}^{2}/\emph{E}_{F}$, which is similar to the behavior of the bound state induced by a single strong nonmagnetic impurity. The agreement in the energy of the bound states for these two types of single impurity reveals a universal feature of the bound state under the strong scattering limit for a chiral $p$-wave superconductor.

\begin{figure}[!t]
\centering
\resizebox{8.1cm}{12.4cm}{\includegraphics{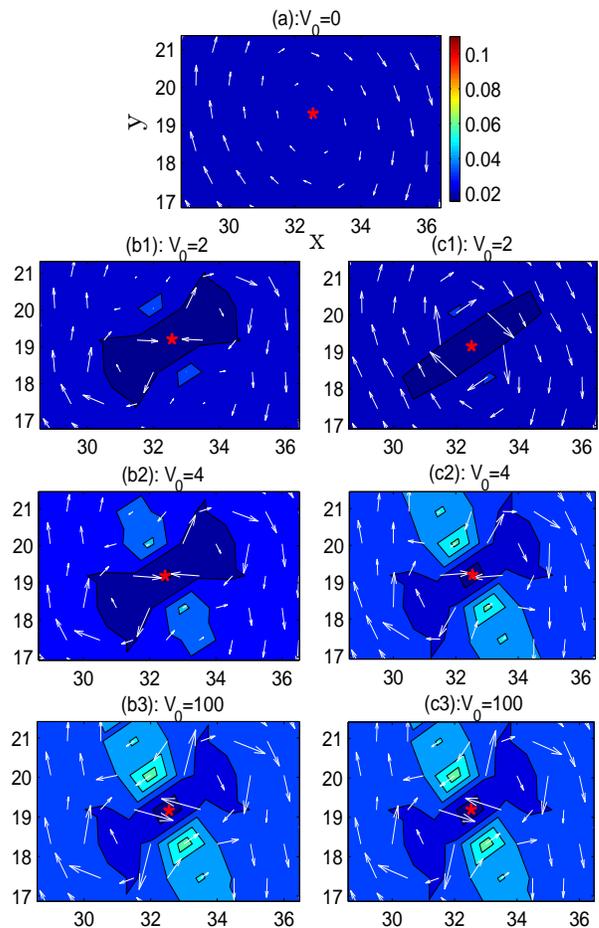}}
\caption{(Color online) Spatial variations of the supercurrent around a single impurity with (a) zero scattering strength and various scattering strengths for (b1)--(b3) a nonmagnetic impurity and (c1)--(c3) a magnetic impurity. The color maps
of the amplitudes of the order parameter correspond to the \emph{$p_{x}- ip_{y}$}-wave. The red asterisk represents the location of the impurity. }
\label{fig:fig61}
\end{figure}

We next discuss the behaviors of the bound states. The values of the order parameter at the site near the nonmagnetic or magnetic impurity are shown in Fig.~\ref{fig:fig2}. Because the order parameter of the \emph{$p_{x}- ip_{y}$} component is negligible compared with that of the \emph{$p_{x}+ ip_{y}$} component in the absence of impurities, the introduction of a nonmagnetic impurity suppresses the pairing order parameter of the \emph{$p_{x}+ ip_{y}$} component and enhances that of the \emph{$p_{x}- ip_{y}$} component. As $\emph{V}_{0}$ increases, the pairing order parameter of the \emph{$p_{x}+ ip_{y}$} (\emph{$p_{x}- ip_{y}$}) component is increasingly suppressed (enhanced). When $\emph{V}_{0}>20$, the values of both pairing components, \emph{$p_{x}\pm ip_{y}$}, tend to saturation. These results are in accordance with the discussion of the LDOS illustrated in Fig.~\ref{fig:fig1}.The pair-breaking effect of the magnetic impurity in Fig.~\ref{fig:fig2}(b) for \emph{$p_{x}+ ip_{y}$}, however, is less ambiguous than that of the nonmagnetic impurity when the strength of the magnetic impurity increases from 0 to 4. When $\emph{V}_{0}$ approaches $4$, the order parameter changes abruptly. This transition is due mainly to resonant-state scattering between spin-up and spin-down species [see Fig.~\ref{fig:fig1}(b)]. As $\emph{V}_{0}$ increases further, the order parameters of the \emph{$p_{x}\pm ip_{y}$} wave for the magnetic impurity tend to have a universal value with unitary scattering. We thus find that changes in the order parameters of the \emph{$p_{x}\pm ip_{y}$} wave correspond to the emergence of the universal impurity-induced bound state. In addition, we note that a single impurity not only induces the universal resonant state but also affects the distribution of the supercurrent and chirality of the order parameters for a chiral $p$-wave superconductor.

Fig.~\ref{fig:fig61} shows the vectors of the supercurrent distribution around the impurity. When the strength of the impurity is set to zero, $V_{0}=0$, the supercurrent has a clockwise structure [see Fig.~\ref{fig:fig61}(a)], in accordance with the chirality of the pairing order parameter of the $p_x+ip_y$ component. As $V_{0}$ increases, the impurity potential affects the local supercurrent distribution. With a further increase in $V_{0}$, the chirality of the local vector of the supercurrent changes. This suggests that the pairing order parameter of the $p_x-ip_y$ component is enhanced by scattering from the impurity. This is consistent with the calculation of the pairing order parameter in Fig.~\ref{fig:fig2}. The amplitude of the \emph{$p_{x}- ip_{y}$} wave can be seen behind the vector-represented supercurrent on the color map of Fig.~\ref{fig:fig61}(b), which clearly shows the two domain structures of the \emph{$p_{x}- ip_{y}$} wave around the nonmagnetic impurity at the strong scattering limit. Furthermore, similar features are obtained for the magnetic impurity, as shown in Fig.~\ref{fig:fig61}(c). These calculations demonstrate universal behaviors of both nonmagnetic and magnetic impurities in a topological superconductor.

\begin{figure}[!t]
\centering
\resizebox{8.5cm}{7.5cm}{\includegraphics{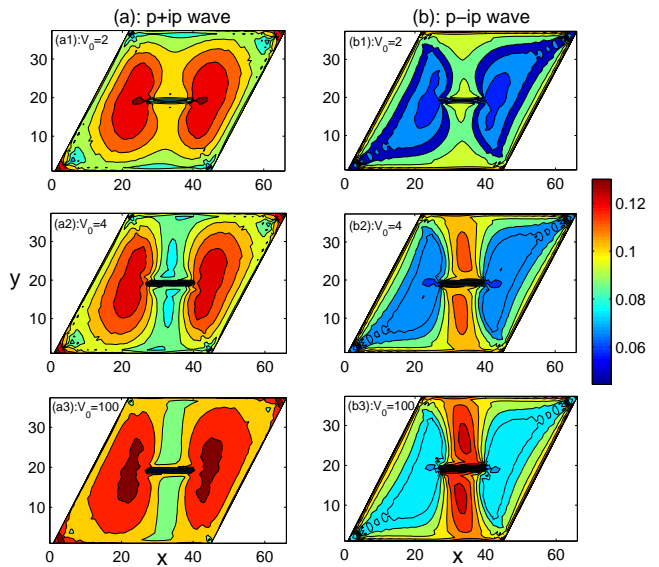}}
\caption{(Color online) Spatial variations of the order parameter around nonmagnetic linear impurities with various scattering strengths $V_{0}$.}
\label{fig:fig3}
\end{figure}

\subsection{Effects of Linear Impurities}

To clearly distinguish the two degenerate components of a chiral $p$-wave superconductor, we examine the effects of linear impurities on a degenerate superconductor because of our finding that a single impurity changes the chirality of the supercurrent around the center of the impurity. Because spontaneous symmetry breaking occurs in the degenerate states, domain walls will emerge, and the system will display a phase-separated structure. In this section, we propose the use of the supercurrent induced by the linear impurities to identify a spin-triplet chiral $p$-wave state to detect the chiral domains. In a chiral superconductor, the pairing potential is generally composed of two degenerate components, \emph{$p_{x}\pm ip_{y}$}. Owing to interaction, the degenerate chiral states favor being broken and forming a domain structure near a row of linear impurities. The spatial chiral domain structures can be resolved by scanning tunneling spectroscopy~\cite{Y.Tanuma}. Nonmagnetic and magnetic linear impurities are discussed separately in detail as follows.

For nonmagnetic linear impurities, the order parameters of \emph{$p_{x}\pm ip_{y}$} are shown in Fig.~\ref{fig:fig3}(a) and (b). Figure~\ref{fig:fig3}(a) shows the dominance of the \emph{$p_{x}+ ip_{y}$} wave throughout the lattice system for $V_{0}=2$. Because the superconducting states along the direction of the linear impurities are sensitive to the potential of the linear impurities, the intensity of the \emph{$p_{x}+ ip_{y}$} wave at the two endpoints of the linear impurities is higher than that on the sides of the linear impurities. However, the intensity of the \emph{$p_{x}- ip_{y}$} wave is so low that it is negligible compared with that of the \emph{$p_{x}+ ip_{y}$} wave [see Fig.~\ref{fig:fig3}(b)]. As $V_{0}$ increases, the higher intensity of the order parameter of the \emph{$p_{x}+ ip_{y}$} wave at the end points of the linear impurities is further enhanced, whereas the weaker intensity of the order parameter of the \emph{$p_{x}+ ip_{y}$} wave along the direction of the linear impurities is further suppressed. The reduced intensity of the order parameter of the \emph{$p_{x}+ ip_{y}$} wave along the direction of the linear impurities is transferred to the side of the order parameter of the \emph{$p_{x}- ip_{y}$} wave, increasing the intensity of the order parameter of the \emph{$p_{x}- ip_{y}$} wave along the direction of the linear impurities. At the limit of the increase in $V_{0}$, the order parameters of the \emph{$p_{x}\pm ip_{y}$} waves both form a stable and universal distribution of the chiral domain structure.

\begin{figure}[!t]
\centering
\resizebox{8.5cm}{7.5cm}{\includegraphics{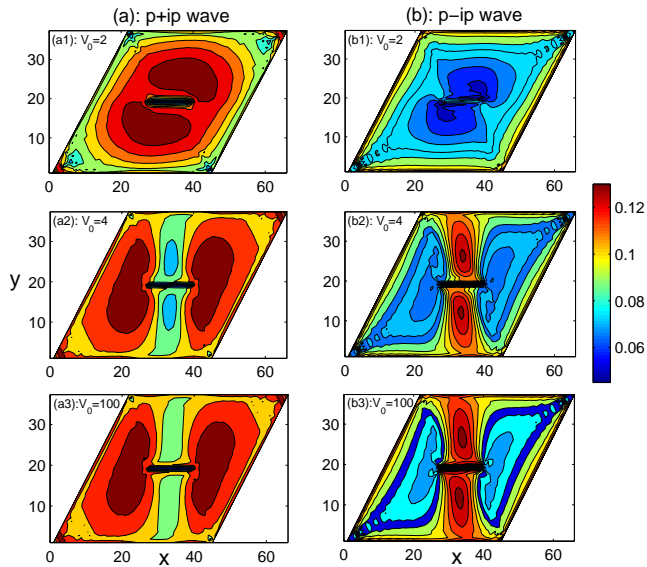}}
\caption{(Color online) Spatial variations of the order parameter around linear magnetic impurities with various scattering strengths $V_{0}$.}
\label{fig:fig5}
\end{figure}

For magnetic linear impurities, the order parameters of the \emph{$p_{x}\pm ip_{y}$} waves are shown in Fig.~\ref{fig:fig5}(a) and (b). At a low strength of $V_{0}=2$ [Fig.~\ref{fig:fig5}(a1)], the dominant order parameter distribution throughout the lattice space comes mainly from the contribution of \emph{$p_{x}+ ip_{y}$}. Unlike the case for nonmagnetic linear impurities shown in Fig.~\ref{fig:fig3}, the intensity of the order parameter at the two endpoints of the linear impurities is much lower than that along the direction of the linear impurities. However, similar to the case for nonmagnetic linear impurities, the strength of the \emph{$p_{x}- ip_{y}$} wave is so low that it can be ignored [see Fig.~\ref{fig:fig5}(b1)]. With increasing $V_{0}$, the intensity transfer between the \emph{$p_{x}\pm ip_{y}$} waves is similar to that for nonmagnetic linear impurities shown in Fig.~\ref{fig:fig3}.

\begin{figure}[!t]
\centering
\resizebox{8.0cm}{8.6cm}{\includegraphics{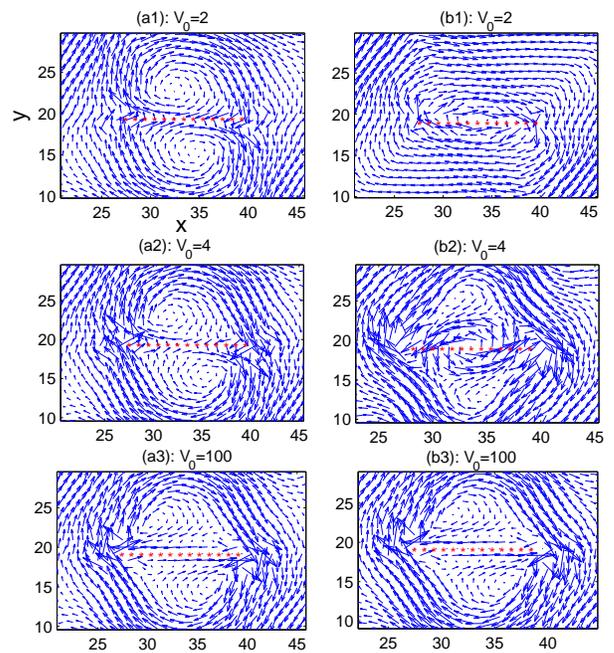}}
\caption{(Color online) Spatial variation of the supercurrent around linear impurities: (a1)--(a3) linear nonmagnetic impurities and (b1)--(b3) linear magnetic impurities with various scattering strengths $V_{0}$.
Red asterisks represent the locations of impurities.}
\label{fig:fig6}
\end{figure}

The numerically obtained supercurrent vector distribution is illustrated in Fig.~\ref{fig:fig6} to clarify the chiral domain structure of the order parameter. Because time-reversal symmetry is preserved for nonmagnetic linear impurities, a two-vortex structure emerges when the linear impurities are inserted [see Fig.~\ref{fig:fig6}(a1)], and the vortex directions are consistent with the chirality of \emph{$p_{x}+ ip_{y}$}. However, because introducing linear magnetic impurities breaks time-reversal symmetry, the enhanced scattering between the two vortices leads to a structure consisting of two invisible vortices [see Fig.~\ref{fig:fig6}(b1)], resulting in a nonuniversal chiral domain structure of the order parameter. As $V_{0}$ increases, the coherence between the two vortices in the nonmagnetic linear impurity system is further suppressed, and the intensity of the vortices on both sides of the linear impurities increases, whereas the enhancement of the strong spin-dependent scattering in the magnetic linear impurity system drives the chirality of the supercurrent structure around the impurities into disorder. Eventually, at the limit of strong $V_{0}$, the nonmagnetic and magnetic linear impurity systems display similar local disordered supercurrent structures, as expected intuitively.

\section{Vortex core states}
\label{sec:splitting}

We now discuss the quasiparticle states in the vortex structure of a chiral \emph{p}-wave superconductor in the presence of an external magnetic field. Importantly, these vortex structures display an intriguing sixfold symmetry consistent with the triangular lattice structure~\cite{hanqiang}. The magnetic flux can generally penetrate the superconductor and form an Abrikosov vortex lattice in a vortex state. For the vortex state on a chiral \emph{p}-wave superconductor, the lowest vortex bound states have zero energy, which is referred to as the Majorana mode, in contrast to the case on an \emph{s}-wave or \emph{d}-wave superconductor~\cite{ccaroli,volovik,ychen1}. The stability of the Majorana mode protected by the symmetries has potential applications in topological quantum computation. We know that the zero-energy (Majorana) mode of the vortex core state on a superconductor is robust against impurity scattering~\cite{scatter,Y.Tanuma} and order parameter perturbations~\cite{perturb}. However, owing to the presence of the degenerate zero-energy mode, intervortex quasiparticle tunneling will destroy such a Majorana mode~\cite{split, split2}. A previous study discussed the splitting of the LDOS peak induced by intervortex scattering in $s$-wave and $d$-wave superconductors in the presence of an external magnetic field~\cite{Donald}. For a chiral \emph{p}-wave superconductor, a previous study also pointed out that the LDOS peaks around the vortex core are split by the presence of antiferromagnetic order~\cite{hanqiang}. Here, we concentrate on the effects of the magnetic strength and the strength of the pairing interaction on the vortex core state in a chiral $p$-wave superconductor.

\begin{figure}[!t]
\centering
\resizebox{8cm}{8.4cm}{\includegraphics{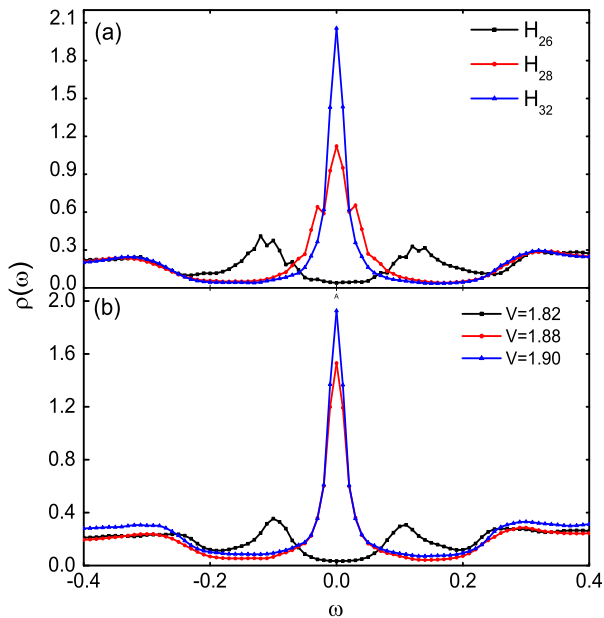}}
\caption{(Color online) Quasiparticle LDOS profiles at the center of vortex cores plotted for (a) various magnetic field strengths and (b) various pairing interaction strengths $V$. In (a), $H_{n}$ represents the magnetic field in areas of $na\times 2na$, with two fixed superconducting flux quanta per system, $2\Phi _{0}$, and the pairing interaction strength $V=2.4$.}
\label{fig:fig11}
\end{figure}

Figure~\ref{fig:fig11}(a) shows the LDOS $\rho_{i}(E)$ at the center of the vortex core, where the strength of the magnetic field changes from $H_{26}$ to $H_{32}$, which corresponds to a change in the size of the magnetic unit cell from $26a\times52a$ and $28a\times 56a$ to $32a\times64a$, with two fixed superconducting flux quanta per system, $2\Phi _{0}$. For a weak external magnetic field, such as $H_{32}$ in Fig.~\ref{fig:fig11}(a), the distance between vortices induced by the magnetic flux is large, the intervortex scattering is weak, and a remarkable zero-energy Majorana mode is visible. As the external magnetic field strengthens, the distance between vortices decreases, and intervortex scattering thus increases. As a result, the intensity of the zero-energy Majorana mode is suppressed. Eventually, the Majorana mode is destroyed and splits into two coherence peaks.

Figure~\ref{fig:fig11}(b) shows the LDOS $\rho_{i}(E)$ in the vortex core for various pairing interaction strengths $V$. For a weak pairing interaction of $V=1.82$, which has low energy compared with the magnetic field, no zero-energy peak is seen. As the pairing interaction strengthens, even beyond the magnetic field on an energy scale, a prominent zero-energy peak of the Majorana mode emerges. The Majorana mode becomes more stable as the pairing interaction strengthens.

\section{Conclusion}
\label{sec:conclusion}

We found a universal bound state induced by a single nonmagnetic impurity or single magnetic impurity at the strong scattering limit for a chiral $p$-wave superconductor according to calculations of the LDOS, order parameters, and supercurrent vector distributions. We further found that different chiral order parameters and the corresponding supercurrents have a uniform distribution around linear impurities. Calculations of the LDOS in the presence of a magnetic field showed that the zero-energy peak intensity in the vortex core can be enhanced dramatically by tuning the strengths of the magnetic field and pairing interaction.

\section*{Acknowledgments}

This work was supported by the National Natural Science Foundation of China (Grant Nos. 11625416 and 11474064), the State Key Programs of China (Grant No. 2016YFA0300504), the Strategic Priority Research Program (B) of the Chinese Academy of Sciences (Grant No. XDB04040300), and the Youth Innovation Promotion Association of the Chinese Academy of Sciences (Grant No. 2016215).

\end{document}